# Variation in Patterns of Metal Accumulation in Thallus Parts of *Lessonia trabeculata* (Laminariales; Phaeophyceae): Implications for Biomonitoring

**Claudio A. Sáez[1,2,3]*, M. Gabriela Lobos[2], Erasmo C. Macaya[4], Doris Oliva[3], Waldo Quiroz[5], Murray T. Brown[1]**

**1** Plymouth University, School of Marine Science & Engineering, Faculty of Science and Technology, Drake Circus, Plymouth, United Kingdom, **2** Universidad de Valparaíso, Departamento de Química y Bioquímica, Facultad de Ciencias. Valparaíso, Chile, **3** Universidad de Valparaíso, Departamento de Biología y Ciencias Ambientales, Facultad de Ciencias. Valparaíso, Chile, **4** Universidad de Concepción, Departamento de Oceanografía, Casilla 160-C, Facultad de Ciencias Naturales y Oceanográficas, Concepción, Chile, **5** Pontificia Universidad Católica de Valparaíso, Laboratorio de Química Analítica y Ambiental, Instituto de Química, Curauma, Valparaíso, Chile

## Abstract

Seaweeds are well known to concentrate metals from seawater and have been employed as monitors of metal pollution in coastal waters and estuaries. However, research showing that various intrinsic and extrinsic factors can influence metal accumulation, raises doubts about the basis for using seaweeds in biomonitoring programmes. The thallus of brown seaweeds of the order Laminariales (kelps) is morphologically complex but there is limited information about the variation in metal accumulation between the different parts, which might result in erroneous conclusions being drawn if not accounted for in the biomonitoring protocol. To assess patterns of individual metals in the differentiated parts of the thallus (blade, stipe, holdfast), concentrations of a wide range of essential and non-essential metals (Fe, Cr, Cu, Zn, Mn, Pb, Cd, Ni and Al) were measured in the kelp *Lessonia trabeculata*. Seaweeds were collected from three sampling stations located at 5, 30 and 60 m from an illegal sewage outfall close to Ventanas, Chile and from a pristine location at Faro Curaumilla. For the majority of metals the highest concentrations in bottom sediment and seaweed samples were found at the site closest to the outfall, with concentrations decreasing with distance from the outfall and at control stations; the exception was Cd, concentrations of which were higher at control stations. The patterns of metal concentrations in different thallus parts were metal specific and independent of sampling station. These results and the available literature suggest that biomonitoring of metals using seaweeds must take account of differences in the accumulation of metals in thallus parts of complex seaweeds.

**Funding:** Source of funding were FONDECYT (Project 11080235) and CONICYT Becas Chile Scholarship (72110557)www.conicyt.cl. The funders had no role in study design, data collection and analysis, decision to publish, or preparation of the manuscript.

**Competing Interests:** The authors have declared that no competing interests exist.

* E-mail: claudio.saez.a@gmail.com

## Introduction

Metals entering estuaries and coastal waters from land-based mining, industrial, agricultural and domestic activities can be readily accumulated by the resident biota, which may display symptoms of toxicity beyond certain threshold concentrations [1,2]. Quantification of metal concentrations in sediments and water provides limited information about bioavailability and therefore analyses of organisms are routinely used to assess the quantity and availability of metals in the environment. The rationale for using organisms to assess levels of environmental pollution and the criteria by which they are selected as biomonitors are well documented [3,4]. Their wide-spread distribution, local abundance, longevity, benthic nature and capacity to accumulate metals to concentrations several thousand times higher than those dissolved in seawater make seaweeds choice candidates. However, because extrinsic and intrinsic factors (e.g. environmental variables affecting growth rates, allometric parameters, interactions between metals) can influence the bioaccumulation of metals, the interpretation of the toxicological significance of thallus burdens may not necessarily be straightforward [5–8].

Brown algae belonging to the orders Fucales and Laminariales have been widely employed in studies on metal pollution in estuarine and coastal ecosystems from temperate latitudes [9–12]. Fucoids and kelps are dominant members of intertidal and sublittoral communities and have vital roles as primary producers and ecosystem bio-engineers [13–16]. They also have the greatest morphological complexity of all seaweeds and there is evidence that the accumulation capacity can differ between thallus parts [10,12,17–19]. This has implications for using seaweeds as biomonitors, as the choice of thallus part sampled could, inadvertently, influence how the data is interpreted [12].

*Lessonia trabeculata* is the most abundant kelp species along the south-east coast of South America, between 18° and 42° S, and constitutes one of the most important components of shallow subtidal ecosystems [20]. Structurally, the thallus can be separated into three main parts: annually renewable blades that are responsible for most nutrient absorption and photosynthetic activity, perennial stipes, providing structural support, and holdfast that anchor the thallus to the substratum [20].

The aim of this study was to assess the patterns of metal concentrations in different metals in the three main components, blades, stipes and holdfasts, of thalli sampled along a pollution gradient emanating from a sewage outfall located in Ventanas, Central Chile, and from a pristine location with similar oceanographic characteristics that acted as a control. The concentrations of the essential metals Fe, Cr, Cu, Zn and Mn and non-essential metals Pb, Cd, Ni and Al were measured. The implications of our findings for using morphologically complex seaweeds in biomonitoring programmes are discussed.

## Materials and Methods

### Study stations

At Ventanas, Central Chile (32°44′32,4′′S; 71°29′56,3′′W), an illegal sewage outfall was identified that constantly discharged wastes within the intertidal zone. The plume from the pipe (30 cm diameter) flowed in a north easterly direction as a consequence of the prevailing wind and water currents from the south west [21]. Three sampling stations, 1V, 2V and 3V, were established along the plume at distances of 5, 30 and 60 m from the outfall, respectively. For comparison, control stations were located at Faro Curaumilla (33°05′38,03′′S; 71°44′08,25′′W), 45 km south of Ventanas (fig. 1), as this area has similar oceanographic and topographic features [22]. Three sampling stations, 1F, 2F and 3F, were established in a northly direction at similar depths to those of the outfall stations and at 5, 30 and 60 m from the shore (low tide mark or chart datum point), respectively. Turbidity of the water at each station was estimated using secchi disks. Three measurements, separated by an interval of 60 min, were taken at each of the six stations between 1100 and 1300 hours. From a boat, the compensation depths (1% of surface incident light) were estimated by the recorded depth of a 25 cm diameter secchi disk multiplied by the constant 2.7 [23].

No permits were required for the described field studies, as both locations are not privately-owned or protected by the Republic of Chile. Locations were public and land-reachable. This study did not involve endangered or protected species.

### Sampling and laboratory pre-treatment

Sampling from Ventanas and Faro Curaumilla was carried out in austral summer, on the 8$^{th}$ and 10$^{th}$ January 2009, respectively. Three independent sediment samples were collected at each sampling station. Sediments were collected with a small plastic shovel, stored in plastic containers and transported to the laboratory where they were dried in an oven (Memmert, model UL30790986) at 60°C to constant weight. Samples were homogenized through a 63 µm nylon sieve. Concurrently with sampling in the outfall zone, 3 samples of effluent were taken at 2 hour intervals for chemical analyses.

At each of the six sampling stations three individual thalli of *Lessonia trabeculata*, holdfasts ranging between 25 to 30 cm diameter with healthy appearance and free from excessive grazing, were collected from the centre of the kelp forest by SCUBA divers. Each thallus was separated into three parts: middle thallus blades, main stipes and central portion of holdfast (see fig. 2). Material was transported to the laboratory in clean plastic bags; seaweeds were rinsed with milli-Q water and a plastic brush was used to remove remainders of sediment and epibiota, dried in an oven at 60°C until constant weight and then pulverized with a titanium blender to powder. Gledhill et al. [24] found that the most effective method for cleaning seaweed material was a solution of 10% ethanol in seawater as evidences from the significant decrease in Fe concentration (compared with washing in milli-Q water). However, the presence of some epibiota on thr surface of *L. trabeculata* cannot be ruled out, as these cleaning methods have been assessed mainly for fucoids. Metal concentrations in three sub-samples of each of the different thallus parts were determined, giving a total of 1458 measurements (9 metals ×3 sub-samples ×3 parts ×3 individual thalli ×3 stations ×2 zones).

### Analytical methods

A protocol of the Environmental Protection Agency was adapted for metal analysis of sediments (EPA, Method 3050B); 0.5 g DW were weighed in an analytical balance and placed in Erlenmeyer flasks. Subsequently, 3 ml of fuming hydrochloric acid (HCl, Merck, 37% G.R. for analyisis), 2 ml hydrofluoric acid (HF, Merck, 48%) and 9 mL $HNO_3$ (Merck, 65% G. R. for analysis) were added for 24 h for pre-digestion. When complete, 6 ml HCl, 2 ml $HNO_3$ and 8 ml of hydrogen peroxide ($H_2O_2$, Merck, 30% G.R. for analysis) were added and digested at 150°C for ~5 h until transparent. The volume was made up to 25 ml with milli-Q water in a volumetric flask. For effluent samples, wastewaters were mixed and 200 ml were concentrated by evaporation to c.5 ml at 95°C. Then, 10 ml of $HNO_3$ were added and samples were digested for 3 hours at 150°C until the digested turned transparent. The final volume was made up to 25 ml with milli-Q water.

The protocol for algal digestion was modified from Burger et al. [12]. Briefly, 1 g dry biomass was placed in an Erlenmeyer flask and pre-digested overnight at room temperature with 20 ml concentrated $HNO_3$ and 4 ml of supra $H_2SO_4$ (Aldrich, supra 96%). Then, 4 ml of $H_2O_2$ were added and digested for 4 to 5 h at 150°C. The solution was cooled to room temperature, 2 ml $HClO_4$ (Riedel- de Haen, 70% G. R. for analysis) then added and the solution re-heated to 150°C for 20–30 min until full digestion. The solution was then filtered (Whatman N°5) and made-up to 25 ml in a volumetric flask with milli-Q water.

The concentrations of Fe, Cr, Cu, Zn, Mn, Pb, Cd, Ni and Al in sediment, wastewaters, and seaweed samples were measured using inductively coupled plasma optical emission spectrometry (ICP-OES, Perkin-Elmer ICP Optima 2000 DV). To validate the analytical procedures, digestion and protocols for metals, measurements were also applied to marine sediment (BCSS-1) and algal (sea lettuce, *Ulva lactuca*, BCR-279) reference materials, the results of which are presented in Table 1.

### Statistical analyses

To assess the requirements of normality and homogeneity of variances, Kolmogorov-Smirnov and Bartlett Test analyses were performed, respectively. One-way ANOVA and post-hoc Tukey test at 95% confidence was estimated for metal concentrations in sediments among sampling stations to identify significant differences. A multivariate analysis of variance (MANOVA) and post-hoc Tukey tests at 95% confidence were performed in order to estimate significant differences between the main factors (zone, stations and tissues) with 9 dependent variables (metals). The error intervals in figures and tables are calculated at 95% confidence ($n = 3$).

Multivariate principal component analysis (PCA), of metal concentrations in blades, stipes and holdfasts from the two locations was also undertaken. This analysis provides a small number of linear combinations of the 9 metal concentrations that describe most of the variability in the data. In addition to the quantitative analysis by the MANOVA, the PCA provides a visual qualitative representation of the data for ease of discerning relationships between metals and sampling stations.

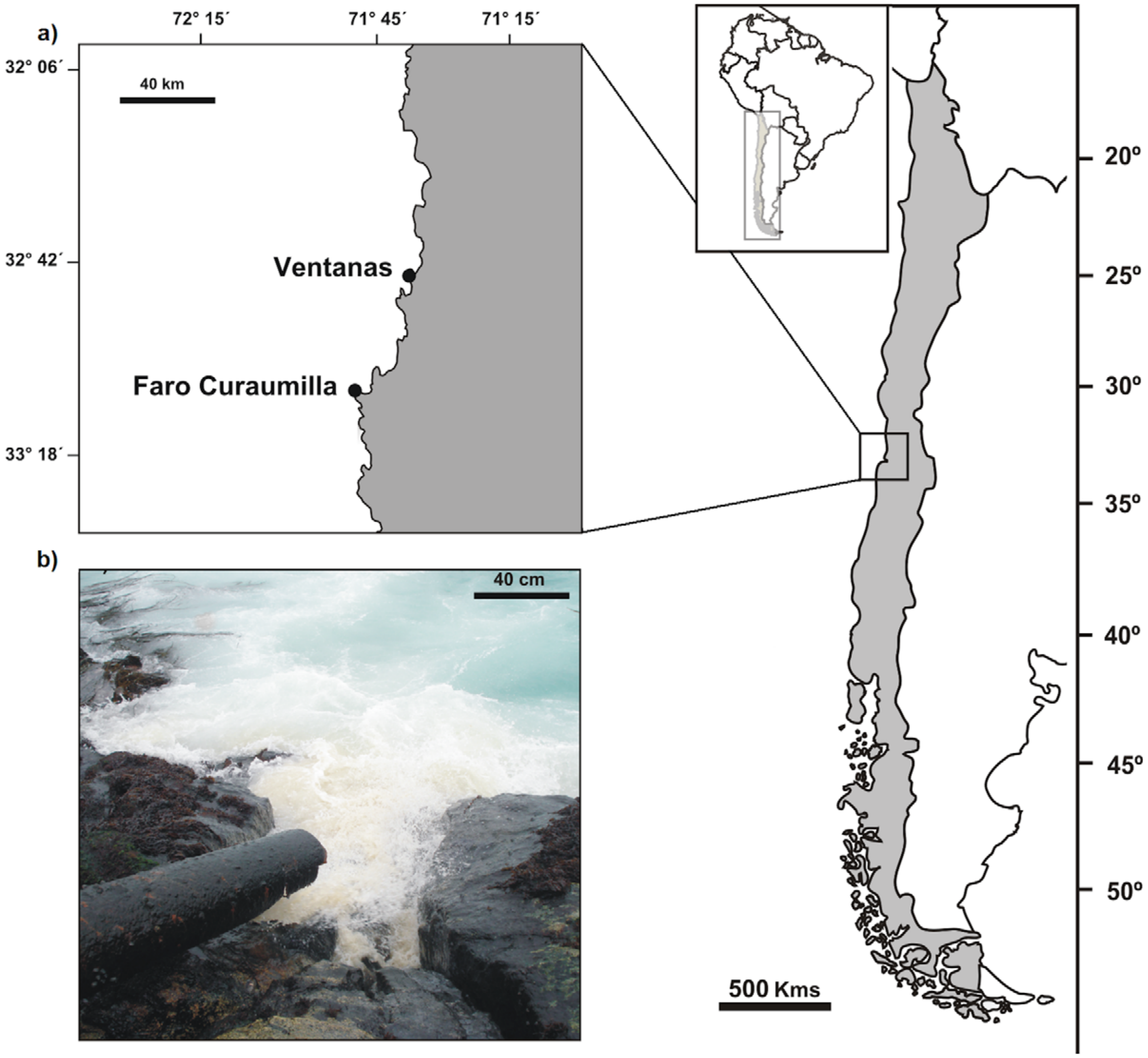

**Figure 1. Map of study zones in Central Chile. a) Enlargement of study zones. b) Photo of the outfall.**

## Results

### Physical measures

Table 2 provides information on the bathymetry and compensation depths of the sampling stations at the two locations (Ventanas and Faro Curaumilla). Despite the same bathymetry at corresponding stations in the two zones, the compensation depths differed significantly ($p<0.001$). The mean compensation depth at the outfall stations was c. 20 m compared with c. 45 m at the control stations.

### Metals in sediments and wastewaters

The metal concentrations in sediments are presented in Table 3. The highest concentrations of Al, Fe, Cr, Cu, Mn, Ni, Pb and Zn were found at the station closest to the outfall (1V), showing significant differences with all the rest stations assessed ($p<0.001$). The concentrations of all these metals decreased significantly with distance from the outflow. For Al, Fe, Mn, Pb and Zn concentrations at 3V were significantly higher than control stations ($p<0.001$) whereas there were no significant differences ($p>0.05$) between 3V and the controls for Cr, Cu and Ni. Due to the detection limit of Cd, this metal was only measurable at stations 2F and 3F.

Metal concentrations measured in the wastewaters and comparisons with the legal limits of permitted concentrations of metals in wastewaters entering the sub-tidal [25], are presented in Table 3. With the exception of Cd (which was below the detection limit of the methodology applied), the concentrations of all other metals measured were higher than the legal permissible limits.

### Metal concentrations in *Lessonia trabeculata*

The results from MANOVA performed using the different metals, sampling stations and algal parts are shown in Table 4. Significant differences ($p<0.00001$) were found for all main effects and 2-way and 3-way interactions. The results from a post-hoc Tukey test allowed for more detailed analysis of these results.

The metal concentrations in *L. trabeculata* are presented in Figure 3, and overall tendencies per algal part assessed in the PCA in Figure 4. For all metals analysed with the exception of Cd, the highest concentrations were found in seaweeds sampled from station 1V with a significant decrease with distance from the outflow pipe. Typically, concentrations at station 3V were significantly higher than those at the control sites. By contrast, Cd concentrations were higher in samples collected in control stations; moreover, in addition to Cd, Zn was out of tendency for blades and stipes while Ni for holdfasts. Some departures from this general trend become apparent when different parts of the seaweed are analysed separately. Concentrations of Fe, Cr, Cu, Pb and Al in blades of seaweeds from station 3V were significantly higher ($p<0.05$) than those sampled from control stations. For Mn there was a trend of decreasing concentrations with distance from the shore at both locations and Cd concentrations were highest at station 3F. There was no obvious pattern for Zn and Ni concentrations were very low at all sampling stations.

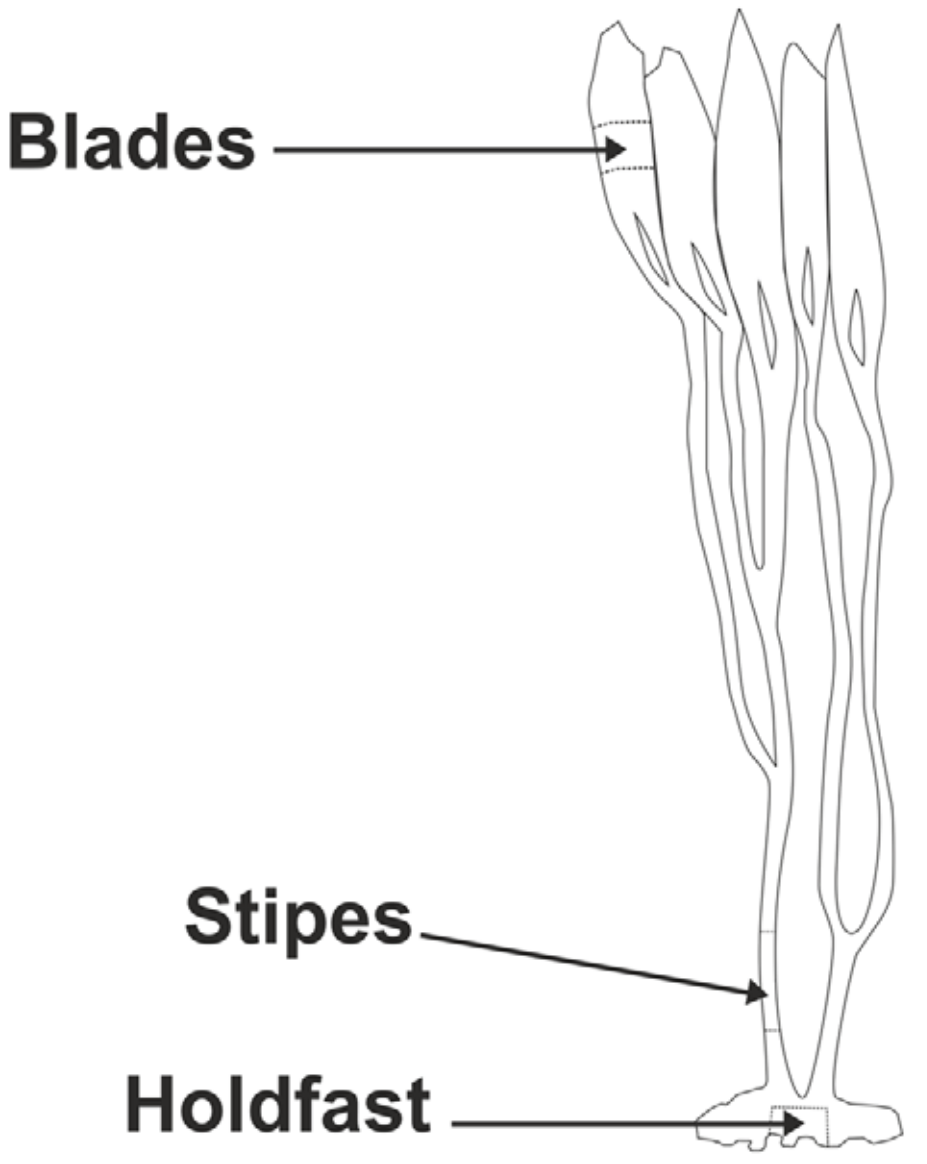

**Figure 2. Diagram of algal thallus parts of *Lessonia trabeculata* sampled for metal analyses.**
doi:10.1371/journal.pone.0050170.g002

For stipes, concentrations of Fe, Cr, Ni, and Al were significantly higher at station 3V than control sites whereas Cu concentrations did not differ significantly between stations. Except for stations 1V and 2V, there were no measurable concentrations of Pb in stipes. The patterns of Mn and Cd concentration in stipes were the same as those of blades and, as with blades, there was no discernible pattern for Zn. The concentrations of Fe, Cr, Mn, Cu, Zn and Al in holdfasts declined with increasing distance from the outfall (1V) and, with the exception of Al, concentrations at station 3V were significantly higher than those at control stations. Concentrations of Pb were detectable in samples from 1V and 2V only, with the highest concentrations at the former station. The trend in Cd was the same as that reported for blades and stipes (Fig. 4).

Table 5 provides a summary of the differences in the concentrations of each metal between the three parts of the seaweed thallus. The highest concentrations of Cu, Pb and Al were present in blades, whereas for Ni it was stipes and for Cd the holdfasts. The lowest concentrations of Cr and Ni were found in blades and for Cd and Al in stipes. For Fe, Mn and Zn there was either no difference or no discernible pattern. With the exception of Zn (blades and stipes) and Ni (holdfasts), there was a significant positive correlation between metal concentrations in the sediment and seaweeds samples (Table 6). Due to the lack of data rather in

**Table 1.** Metal measures in certified material BCSS-1 (marine sediment) and BCR-279 (sea lettuce, *Ulva lactuca*). Confidence intervals at 95% (n = 3).

| | $Al_2O_3$ | $Fe_2O_3$ | Cu | Cr | Cd | Mn | Ni | Pb | Zn |
|---|---|---|---|---|---|---|---|---|---|
| found value (mg $Kg^{-1}$) | 11.0±1.0[a] | 5.00±0.40[a] | 17.0±0.8 | 120±8 | 0.25±0.04 | 214±15 | 56.0±3.0 | 24.0±2.0 | 120±8 |
| certified value (mg $Kg^{-1}$) | 11.8±0.4[a] | 4.70±0.14[a] | 18.5±2.7 | 123±14 | 0.32±0.06 | 229±15 | 55.3±3.6 | 22.7±3.4 | 119±12 |
| | **Al** | **Fe** | **Cu** | **Cr** | **Cd** | **Mn** | **Ni** | **Pb** | **Zn** |
| found value (mg $Kg^{-1}$) | 2.0±0.2 (×103) | 2.5±0.2 (×103) | 12.8±0.4 | 11.0±0.7 | 0.28±0.03 | 1.98±0.12 (×103) | 15.6±0.5 | 13.1±0.5 | 52±2 |
| certified value (mg $Kg^{-1}$) | 1.90±0.05[b] (×103) | 2.4±0.1[b] (×103) | 13.1±0.4 | 10.7±0.9[b] | 0.27±0.02 | 2.09±0.08[b] (×103) | 15.9±0.4[b] | 13.5±0.4 | 51.3±1.2 |

[a]concentration in percentage w/w.
[b]Indicative Values.
doi:10.1371/journal.pone.0050170.t001

**Table 2.** Compensation depths and bathymetry in sampling stations from the outfall and control zones.

| Sampling stations | Compensation depth (m) | Bathymetry (m) |
|---|---|---|
| 1V | 19.4±0.2 | 14 |
| 2V | 20.2±0.2 | 17 |
| 3V | 21.2±0.2 | 20 |
| 1F | 43.2±2.7 | 14 |
| 2F | 44.1±4.1 | 17 |
| 3F | 45.9±2.7 | 20 |

Sampling stations located at 5,30 and 60 m from the sewage pipe at the outfall zone, named 1V, 2V, and 3V, respectively; Same sampling design at control zone, sediment samples taken at 5, 30, and 60 m from the shore (reference point, no outfall), named 1F, 2F, and 3F, respectively. Compensation depths correspond to bathymetry multiplied by the constant 2.7 [23]. Confidence intervals at 95% (n = 3).

sediments or algae material, for Cd in all the tallus parts assessed and for Pb in stipes and holdfasts, correlation analyses were not possible.

## Discussion

Concentrations of Fe, Al, Cu, Cr, Mn, Ni, Pb and Zn in most thallus parts of *L. trabeculata* collected from the metal impacted stations of Ventanas were significantly higher than in material collected from the pristine stations of Faro Curaumilla, with concentrations of metals associated with thalli decreasing with distance from the sewage outflow pipe. The exception to this general pattern was Cd, the concentrations of which were higher in algal parts from sampling stations of the control zone. There are several possible interpretations for these findings. For example, the presence of high concentrations of metals in the outfall zone might have impeded the accumulation of Cd. Simultaneous exposure of several metals can modify sorption and accumulation of certain metals, including Cd. For instance, Zhou et al. [26] found that the sorption capacity of *Laminaria japonica* for Cd decreased from 97 mg $L^{-1}$ when exposed to 400 mg $L^{-1}$ to 39 mg $L^{-1}$ in the presence of Cu cations. However, considering that highest concentrations of Cd found in sediments were from stations of the control zone and were undetectable at the outfall it is most likely that the upwelling characteristics of Faro Curaumilla, together with the lack of competition from other metals for binding sites, are responsible for the higher than expected levels of Cd in *L. trabeculata*. Faro Curaumilla is located within a recognized upwelling system [22] and it is known that Cd has a close affinity with phosphates and nitrates in the marine environment [27–31]. Complexed Cd is removed from the upper layer of the water column by planktonic organisms and is incorporated into their biomass. After sedimentation of the biogenic material, organic oxidation returns nutrients and Cd to the water column which can then be transported to coastal zones by upwelling processes [32] and becomes bioavailable to seaweeds.

The most interesting results of this study is the lack of consistency in the patterns of accumulation for the different metals between thallus parts of *L. trabeculata*. Brown algae are known to accumulate metals to several orders of magnitude higher than concentrations dissolved in seawater, and they integrate short-term temporal fluctuation in concentrations. For these reasons they have been often used as monitors of metal pollution

**Table 3.** Metal concentrations in sediments and measured in the outfall wastewaters.

| Sediments in stations | Al | Fe | Cd | Cr | Cu | Mn | Ni | Pb | Zn |
|---|---|---|---|---|---|---|---|---|---|
| 1V | 131±15[a] | 464±18 | <DL | 23±3 | 68±3 | 262±13 | 4,0±0,4 | 6,8±0,3 | 51±6 |
| 2V | 84±3[a] | 251±38 | <DL | 9,6±0,5 | 58±4 | 201±10 | 3,1±0,1 | 3,5±0,4 | 28±4 |
| 3V | 65±5[a] | 111±7 | <DL | 5,6±0,3 | 18±2 | 109±6 | 2,0±0,1 | 1,9±0,1 | 20±1 |
| 1F | 45±4[a] | 51±4 | <DL | 4,8±0,2 | 16±1 | 67±8 | 1,8±0,3 | 1,0±0,1 | 14±2 |
| 2F | 48±5[a] | 54±3 | 0,45±0,01 | 5,0±0,2 | 14±1 | 70±10 | 1,9±0,1 | 1,1±0,1 | 12±2 |
| 3F | 43±4[a] | 50±1 | 0,45±0,04 | 4,7±0,3 | 14±1 | 70±2 | 1,9±0,2 | 1,1±0,1 | 15±2 |
| **Wastewaters** | **Al** | **Fe** | **Cd** | **Cr** | **Cu** | **Mn** | **Ni** | **Pb** | **Zn** |
| Sewage | 2269±221 | 242±11 | <DL | 4,9±0,1 | 240±7 | 66±4 | 5,1±0,2 | 3,6±0,3 | 132±15 |
| Permitted concentrationb | 1.0 | 10.0 | 0.02 | 2.5 | 1.0 | 2.0 | 2.0 | 0.2 | 5.0 |

[a]Concentrations in mg $g^{-1}$.
[b]According to SEGEGOB (2001), local law in regard to the permitted limits of metal concentrations in wastewaters discarded to the subtidal.
<DL: Below the detection limit.
Last row describes permitted metal concentrations in sewage by local law. Sediment samples taken at 5,30 and 60 m from the sewage pipe at the outfall zone, named 1V, 2V, and 3V, respectively; Same sampling design at control zone, sediment samples taken at 5, 30, and 60 m from the shore (reference point, no outfall), named 1F, 2F, and 3F, respectively. Metal concentrations in sediments and water in µg $g^{-1}$ and µg $L^{-1}$, respectively. Confidence intervals at 95% (n = 3).

**Table 4.** Multivariate analysis of variance (MANOVA) at 95% confidence showing interactions between main factors (zones, sampling stations, and algal parts) and 9 dependent variables assessed (metals).

| Effect | Value | F | Hypothesis df | Error df | Significance |
|---|---|---|---|---|---|
| Intercept | 0.000 | 32150.869 | 9.000 | 28.000 | 0.000 |
| Zone | 1.000 | a | 0.000 | 32.000 | |
| Sation | 0.000 | 94.725 | 36.000 | 106.666 | 0.000 |
| Algal part | 0.000 | 1471.306 | 18.000 | 56.000 | 0.000 |
| Zone*Station | 1.000 | a | 0.000 | 32.000 | |
| Zone*Algal part | 1.000 | a | 0.000 | 32.000 | |
| Sation*Algal part | 0.000 | 63.557 | 72.000 | 177.897 | 0.000 |
| Zone*Station*Algal part | 1.000 | a | 0.000 | 32.000 | |

a. Exact statistic due to factor Zone has 2 levels, the outfall and control zones.
Wilks' lambda distribution is shown for multivariate hypothesis testing.

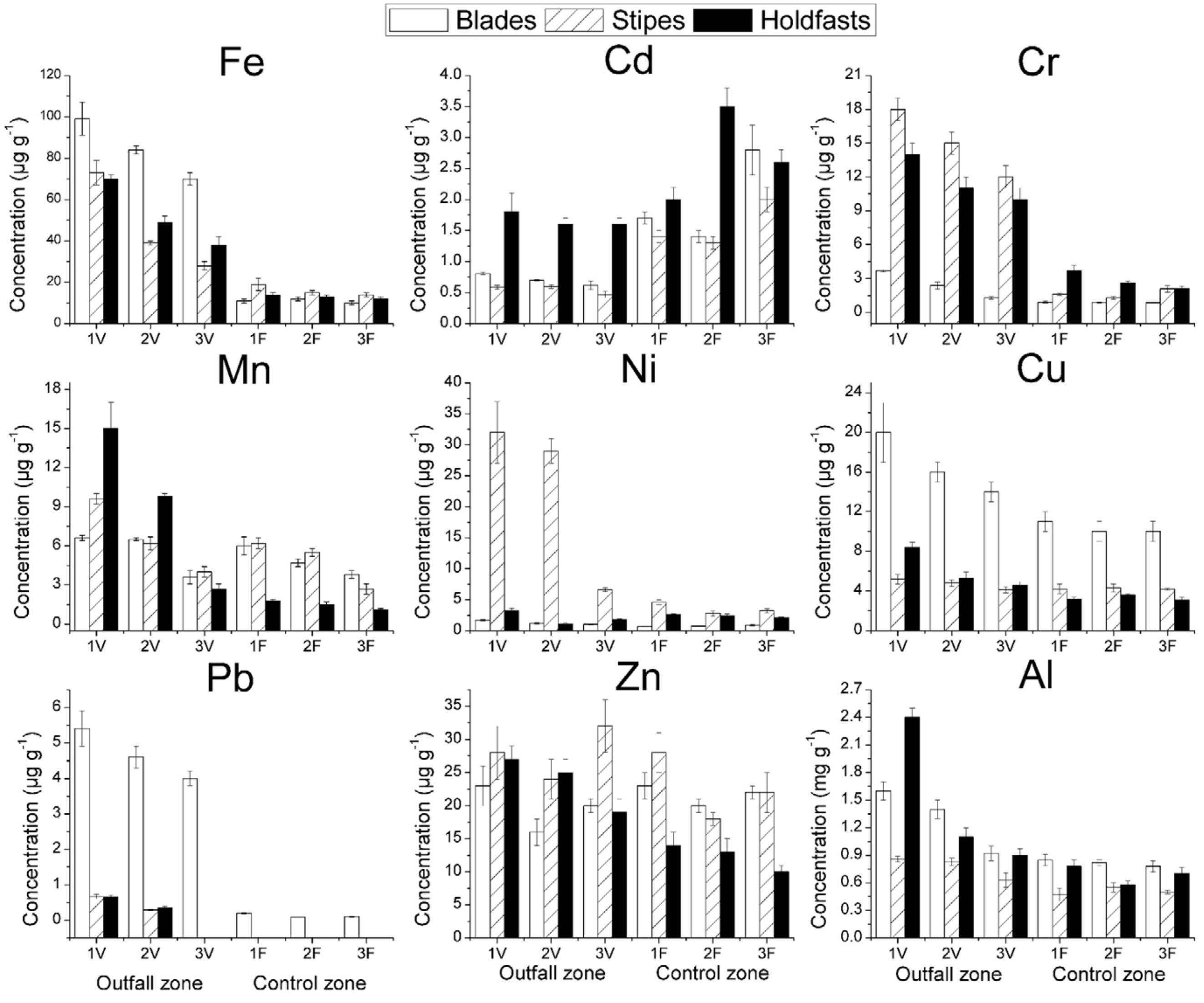

**Figure 3. Metal concentrations in different thallus parts of *L. trabeculata* collected from different sampling stations.** Sampling stations located at 5,30 and 60 m from the sewage pipe at the outfall zone, named 1V, 2V, and 3V, respectively; Same sampling design at control zone, sediment samples taken at 5, 30, and 60 m from the shore (reference point, no outfall), named 1F, 2F, and 3F, respectively. Sampling carried out during austral summer (January) of 2009. Intervals at 95% confidence (n = 3).

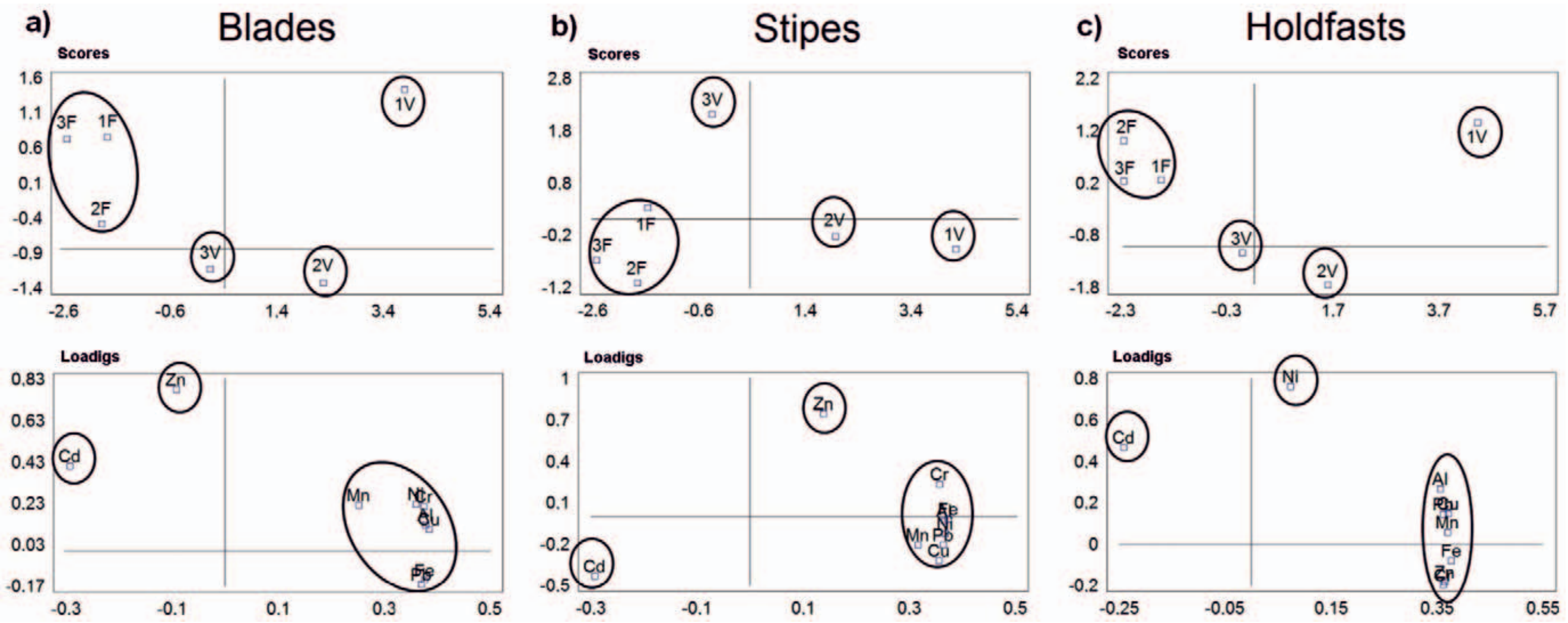

**Figure 4. Principal component analysis (PCA) of metals measured in blades, stipes and holdfasts of *L. trabeculata* from the different sampling stations.** Sampling stations located at 5, 30 and 60 m from the sewage pipe at the outfall zone, named 1V, 2V, and 3V, respectively; Same sampling design at control zone, sediment samples taken at 5, 30, and 60 m from the shore (reference point, no outfall), named 1F, 2F, and 3F, respectively.
doi:10.1371/journal.pone.0050170.g004

[33]. The accumulation and uptake of metals by brown algae are affected by both environmental factors that influence metabolism and growth (e.g. light, temperature) and alter metal speciation, and also by intrinsic factors (e.g. age, allometry, tissue type) [33]. For example, it has been observed that both essential and non-essential metals can have an influence on accumulation and uptake, for example by increasing Cu accumulation during growth in *Ecklonia radiata* [18] or by not having influence on growing tissue, in case of the non- essential Cd in *Laminaria saccharina* [34]. Several studies have identified the holdfast as the structure that accumulates the highest concentration of metals [10,12,35–41], while others found no or different patterns within thalli [10,42,43]. For example, Stengel et al. [10] reported higher concentrations of Cu, Fe and Mn in holdfasts of *Laminaria digitata*, compared to other parts of the thallus at both polluted and pristine locations. They attributed to time dependent accumulation in perennial holdfasts and to their closer proximity to leached ions that might be released from bottom sediments. However, in *Fucus vesiculosus* no such differences in metal concentrations along the thallus were observed. In this study on *L. trabeculata*, concentrations of Cu and Pb were highest in blades, Ni in stipes, Cd in holdfasts and Cr in holdfasts and stipes. We could not observe signs of saturation, even in the point where concentrations of the majority of metals in sediments were highest (sampling site closest to the outfall). Thus, the reported intra-individual variation in metal burdens between tissue structures appears inconsistent and seemingly dependent on the metal and species of seaweed. The reasons for the different patterns of accumulation are not understood but are likely to be related to a number of physico-chemical and biological factors.

It is generally agreed that the first line of defence against metal toxicity is the cell wall. [33,44]. The main components of cell walls in brown algae (Phaeophyta) are cellulose, alginate, fucoidan and mucilage [45]. The alginate content of different brown algae can vary between 10 and 40% of the dry weight [46], and is most important macromolecule involved in metal chelation. The carboxylic groups between segments of polymers forming alginate

**Table 5.** Tendencies from MANOVA and *posteriori* Tukey test.

| Metal | Alga parts distribution |
|---|---|
| Al | blades > holdfasts > stipes |
| Cu | blades > stipes ~ holdfasts |
| Pb | blades > stipes ~ holdfasts |
| Ni | stipes > holdfasts > blades |
| Cd | holdfasts > blades > stipes |
| Cr | holdfasts ~ stipes > blades |
| Fe | blades ~ stipes ~ holdfasts |
| Mn | blades ~ stipes ~ holdfasts |
| Zn | blades ~ stipes ~ holdfasts |

In algal parts distribution signs > and < represent higher and lower than, respectively, when significant differences were found ($p<0.05$), while ~ means no significant differences ($p>0.05$) or no pattern recognized between metal concentrations among algal parts.
doi:10.1371/journal.pone.0050170.t005

**Table 6.** Correlations (R values) of metal concentrations in sediments and thallus parts of *L. trabeculata*.

| Metals | Blades | Stipes | Holdfasts |
|---|---|---|---|
| **Al** | 0.9587 | 0.9146 | 0.9639 |
| **Fe** | 0.8681 | 0.9907 | 0.9553 |
| **Cd** | ID | ID | ID |
| **Cr** | 0.9618 | 0.7721 | 0.7846 |
| **Cu** | 0.9272 | 0.9662 | 0.8975 |
| **Mn** | 0.7843 | 0.7692 | 0.9919 |
| **Ni** | 0.9616 | 0.9646 | 0.2164 |
| **Pb** | 0.8302 | ID | ID |
| **Zn** | 0.1005 | 0.3564 | 0.8682 |

ID: Insufficient data to perform representative correlations.
doi:10.1371/journal.pone.0050170.t006

are key targets for metal ions such as $Pb^{2+}$, $Cu^{2+}$, $Cd^{2+}$, and $Zn^{2+}$ [47,48]. Alginic acid is composed of linear polysaccharides containing mannuronic (M) and guluronic (G) acid residues, organized as (–M–)n, (–G–)n, and (–MG–)n [49]. Changes in the molecular conformation of alginic acid are responsible for differences in their affinity for metals. For example, Haug [48] found that the lower M/G ratio the binding capacity of alginates for the divalent metal ions Cu, Zn, Pb, and Cd, increased. As well as different algal species varying in quantity and chemical composition of alginate, different structural components of the brown seaweed thallus can have different forms and quantities of alginate, which are influenced by the environmental conditions in which the seaweeds are growing [50]. For example, Chandia et al. [51] observed no systematic differences in M/G ratios between holdfasts, stipes and blades in *L. trabeculata*, from specific sampling sites; however, when comparing individuals from different locations, significant differences in M/G ratios were observed between thallus parts. In addition, Venegas et al. [52] found that M/G ratios were lower in blades of *L. trabeculata* collected from a beach exposed to strong currents, in comparison to blades collected from a protected bay. The information suggests that M/G ratios can vary between *L. trabeculata* populations exposed to different environmental factors and the development of divergent ecotypes are likely to occur under different environmental conditions which could influence responses to metal exposure. Further research on the influence of environmental change on alginate content and composition of thallus parts would contribute to our understanding of why metal bioaccumulation differs between thallus parts of *L. trabeculata*.

Seaweeds exude dissolved organic compounds from their cells that chelate metals in the water column, thereby altering their bioavailability in the surrounding water and reducing accumulation and uptake into cells [53]; Of these phenolic compounds are released in large quantities by brown algae and are important chelators of metals in the environment [54–59]. Phenolic compounds are associated with physodes in cells of brown seaweeds. Their content varies between brown algae species and also in different thallus parts of an individual [60,61]. They can be variously induced by, for example, herbivory, nitrogen availability as well as exposure to metals, all of which could influence the metal accumulation capacity of specific algal tissues [57–59].

Metals transported across the cell membrane and entering cells are potentially toxic to cell functions. In order to prevent this occurring metals must be effectively sequestered by, for example, polyphenols, glutathione, phytochelatins, metallothioneins and then either stored in organelles such as vacuoles and physodes or subsequently released to the environment [62]. The method we used to measure concentrations of metals associated with seaweed tissues does not allow for discrimination between externally bound metals to cell surfaces and the internalised fractions. However, previous studies have shown that the proportion of extra-cellular to intra-cellular bound metal differs for different metals, between seaweed species, and can vary seasonally [63].

The behaviour of Zn was another interesting point of this investigation, while holdfasts showed clear influence of the outfall pipe and therefore environmental representativeness, structures such as blades and stipes did not evidence differences that could be attributed to surroundings levels of Zn. Concentration of metals Fe, Cd, Cr, Cu and Al showed environmental patterns in all the algal parts assessed. Tendencies were observed in the rest of the metal measured, however, they were clear only for certain thallus parts; for Mn and Pb blades were more representative, while for Ni and Zn, stipes and holdfasts, respectively, were the most representative thallus parts. Even though it might be argued that total metal concentrations in sediments do not represent the bioavailable fraction for uptake and accumulation, the strong relationship between total metal concentrations in sediments and in the different algal parts of *L. trabeculata* suggests that also the bioavailable fraction of these metals followed a decay from the point of sewage discharge, and lower concentrations in stations from the control zone (except Cd). In our case, the information gathered on metal burdens in *L. trabeculata* suggests that the outfall zone in Ventanas is polluted by action of the sewage pipe. Taking into account that the wastewaters released from the outfall far exceeded the permitted concentrations of metals by the local legislation [25] and, moreover, the records available on the negative effects of this outfall on surrounding biological communities [16], indicate that *L. trabeculata* metal concentrations can reliably constitute a pollution discriminatory tool for biomonitoring the coastal rocky shore.

Our investigation can be considered to be a more complex example of 'passive' biomonitoring (PBM), in which metal measurements have been taken from native individuals that had recruited and grown in the ecosystems assessed. Several authors [49,64–66] have suggested that 'active' biomonitoring (ABM) is a better option to study environmental pollution. This approach employs specific and responsive organisms that have been cultured in the laboratory under controlled conditions and then introduced into locations of environmental interest for a defined period of time. ABM provides several advantages including: using individuals of similar age and state of development and physiology, responses can be observed between introduced and native individuals, experimental locations can be selected despite the absence of the biomonitoring species. However, not all species are necessarily appropriate for this approach. In the case of *L. trabeculata*, and in the majority of complex seaweeds, as well as the variability in metal accumulation between different thallus parts, their culture and growth in the laboratory is demanding of time and logistics. While it would be possible to culture and deploy young kelp sporophytes, it may be easier to use morphologically less complex, perennial, species of seaweeds for ABM, albeit only for short-term evaluations.

## Conclusions

Metal concentrations varied between blades, stipes and holdfasts of *L. trabeculata*. In most cases patterns of accumulation were independent of the sampling station assessed (pristine or polluted). The results indicate that the concentrations of metals associated with the thalli of *L. trabeculata* can discriminate between metal-polluted and pristine locations. However, the information must be treated carefully, as metal accumulation can vary significantly among thallus parts, and that this variation can be metal specific. Our results suggest that different thallus parts of *L. trabeculata* could be sampled for the assessment of particular metals, but no one part is suitable for all metals. Given the widespread use of seaweeds as biomonitors of metal pollution, it is important to consider the outcome of our investigation when other morphologically complex seaweeds are included in evaluating levels of environmental pollution. The known innumerable environmental and species specific intrinsic factors that can influence metal uptake and accumulation bring uncertainties that need to be addressed prior to the selection of candidate species for biomonitoring programmes in order to prevent false diagnosis.

## Acknowledgments

The cooperation of René Sáez, Danilo Gutierrez, Claudio Galarce, Felipe Santander and Patricia Díaz during the different stages of this research is greatly appreciated. We are grateful of Dr. Miguel Franco of Plymouth University for his valuable statistical advice. We also thank the comments of two anonymous reviewers that helped improve this manuscript.

## Author Contributions

Conceived and designed the experiments: CAS MGL DO. Performed the experiments: CAS MGL. Analyzed the data: CAS MGL LECM DO WQ MTB. Contributed reagents/materials/analysis tools: CAS MGL DO WQ. Wrote the paper: CAS MGL LECM DO WQ MTB.